\documentclass[12pt]{article}
\usepackage{latexsym}
\usepackage{amssymb}
\usepackage[english]{babel}
\usepackage{amsmath}
\title{The convolution model of unstable particles}
\author{V.I. Kuksa}
\date{Institute of Physics, Rostov State University,\\
 pr. Stachki 194, Rostov-on-Don, 344090 Russia,\\
 E-mail address: kuksa@list.ru}
\begin{document}

\maketitle
\begin{abstract}

Quantum field model of unstable particles with random mass is
suggested to describe the finite-width effects in decay rate.
Within the framework of this model we derive the convolution
formula for a width of the channels with unstable particle in a
final state. The distribution function of random  mass is
considered for unstable particles of arbitrary type.

PACS number(s): 11.10St, 130000


\end{abstract}

\pagenumbering{arabic} \setcounter{page}{1}

\section{Introduction}

Quantum field description of the unstable particles (UP) with a
large width runs into some problems, which are under considerable
discussions \cite{1}. These problems have both the conceptual and
technological status and arise due to UP lie somewhat outside the
traditional formulation of quantum field theory \cite{2}. Unstable
field can not be treated as asymptotic state and perturbative
approach is unfit in the resonance neighborhood. This conceptual
problems is connected with methodological difficulties, such as
ambiguity in definition of mass and width. Therefore, the new
quantum field approach \cite{2} (Bohm et al), phenomenological
models \cite{3} and effective theories of UP \cite{4} are actual
now.

Convolution method \cite{5} is convenient and clear
phenomenological way to evaluate the instability or finite-width
effects (FWE). This method describes FWE in the processes of type
$\Phi\rightarrow\phi_1\phi\rightarrow\phi_1\phi_2\phi_3 ...$,
where $\phi$ is UP with a large width. The intermediate unstable
state $\phi$ is simulated by the final state $\phi$ in the decay
$\Phi\rightarrow\phi_1\phi$ with invariant mass, described by
Breit-Wigner-like (Lorentzian) distribution function. The
phenomenological expression for a decay rate has convolution form
\cite {5}:
\begin{equation}\label{E:1}
 \Gamma(\Phi\rightarrow\phi_1\phi) = \int_{q^2_1}^{q^2_2} \Gamma(\Phi\rightarrow
 \phi_1\phi(q))\rho(q)dq^2\,,
\end{equation}
where $\rho(q)=M\Gamma_{\phi}(q)/\pi|P(q)|^2$. In Eq.(\ref{E:1})
$\rho(q)$ is probability density of invariant mass distribution,
$P(q)=q^2-M^2+iM\Gamma_{\phi}(q)$ \cite{5}(Altarelli et al),
$\Gamma(\Phi\rightarrow\phi_1\phi(q))$ and $\Gamma_{\phi}(q)$ are
partial width of $\Phi$ and total width of $\phi$ in the stable
particle approximation, when $m^2_{\phi}=q^2$. The formula for a
decay rate, which has a close analogy to the Eq.(\ref{E:1}), was
applied first to describe FWE in $B$ and $\Lambda$ decay channels
with $\rho(770)$ and $a_1 (1260)$ in the final states, which have
large total widths \cite{6}. It was shown that the contribution of
FWE to the decay rates of these channels are large (20-30 \%) and
that the account of it significantly improves a conformity of
experimental data and theoretical predictions. Analogous results
were obtained in Ref. \cite{3} for the dominant decay channels of
$\Phi(1020)$, $\rho(770)$ and $K^*(892)$. The decay rates of the
near-threshold decay channels $t\rightarrow WZb, cWW, cZZ$ were
calculated with help of convolution formula (CF) in Ref. \cite{5}.
It was shown in these works, that the FWE play a significant role
in the near-threshold processes.

The convolution formula (\ref{E:1}) was derived by direct
calculation from the decay-chain method in Ref. \cite{7}, where in
analogy with inclusive processes the contribution of all decay
channels of UP is described by function
$\rho(q^2)=q\Gamma(q)/|P(q)|^2$. The essential elements of this
derivation for vector and spinor UP was the expressions
$\eta_{mn}=-g_{mn}+q_m q_n/q^2$ and $\hat{\eta}=\hat{q}+q$ as
numerators of vector and spinor propagators
($\hat{q}=q_i\gamma^i$). The convolution formula was derived for
the decay chain $t\rightarrow bW\rightarrow bf_if_j$ in the limit
of massless fermions $f$ \cite{5}(Galderon and Lopez-Castro).
Quantitative analysis of convolution and decay-chain calculations
of the $t\rightarrow WZb$ decay rate was fulfilled in Ref.
\cite{5} (Altarelli et al). The formula for a decay rate, which is
in close analogy with (\ref{E:1}), was received in Ref. \cite{3}
for the scalar UP within the framework of the "random mass" model.
An UP is described in this model as quantum field with a "smeared"
(fuzzy) random mass in accordance to uncertainty principle for
energy and lifetime of unstable quantum system \cite{8}. The FWE
is connected with fundamental principle, which gives the relation
$\delta m*\tau \approx1$, that is $\delta m \approx \Gamma$ in the
rest frame of reference ($\delta E=\delta m$, $c=\hbar
=1$)\cite{3}. So, uncertainty principle leads to the
interpretation of kinematic value $q^2$ in Eq.(\ref{E:1}) as
random mass square. Thus, the intermediate states of UP, which are
traditionally defined as virtual, in the neighborhood of $q^2=M^2$
are not differ from real ones in accordance to uncertainty
principle. This interpretation is connected with a smearing of
mass shell and with above mentioned definition of $\eta_{mn}$ and
$\hat{\eta}$, which are proportional to the polarization matrix
for vector and spinor UP (see section 3). As was noted in
\cite{7}, this proportionality leads to the factorization of
expression for width in decay-chain method, and, as consequence,
to the CF (\ref{E:1}). Thus, the suggested model is theoretical
basis of the convolution method on the ground of uncertainty
principle.

In this paper we consider the generalization of the model
\cite{3}, which includes vector and spinor fields. Within the
framework of this generalized model the CF is derived for UP of
arbitrary type. To determine the probability density $\rho(m)$,
which is an analogue of $\rho(q)$ in Eq.(\ref{E:1}), we put a
connection between the model and effective theory of UP with
modified propagators, used in Ref. \cite{7}. It was shown that
this connection leads to Lorentzian probability density $\rho(m)$,
which was commonly used in convolution method. Suggested model is
applicable to the decay processes of type $\Phi\rightarrow
\phi_1\phi(q)$ and gives the convolution formula (\ref{E:1}) for
UP of arbitrary type.

\section{The model of unstable particles with a random mass}

The effect of mass smearing is described by the wave packet with
some weight function $\omega(\mu)$, where $\mu$ is random mass
parameter \cite{3}. The model field function, which simulates UP
in the initial, final or intermediate states, is represented by
the expression:
\begin{equation}\label{E:2}
 \Phi_{\alpha}(x)=\int \Phi_{\alpha}(x,\mu)\omega(\mu)d\mu\,.
\end{equation}
In Eq.(\ref{E:2}) $\Phi_{\alpha}(x,\mu)$ are the components of
field function, which are determined in the usual way when
$m^2=\mu$ is fixed (stable particle approximation). The limits of
integration will be defined in the sections 3 and 4.

The model Lagrangian, which determines "free" unstable field
$\Phi(x)$, has the convolution form:
\begin{equation}\label{E:3}
 L(\Phi(x))=\int L(\Phi(x,\mu))|\omega(\mu)|^2\,d\mu\,.
\end{equation}
In Eq.(\ref{E:3}) $L(\Phi(x,\mu))$ is standard Lagrangian, which
describes model "free" field $\Phi(x,\mu)$ with fixed mass
$m^2=\mu$.

From Eq.(\ref{E:3}) and prescription
$\partial\Phi(x,\mu)/\partial\Phi(x,\mu^{'})=\delta(\mu-\mu^{'})$
it follows Klein-Gordon equation for each spectral component:
\begin{equation}\label{E:4}
 (\square-\mu)\Phi_{\alpha}(x,\mu)=0.
\end{equation}
As a result we have standard momentum representation of field
function for fixed mass parameter $\mu$:
\begin{equation}\label{E:5}
 \Phi_{\alpha}(x,\mu)=\frac{1}{(2\pi)^{3/2}}
 \int\Phi_{\alpha}(k,\mu)\delta(k^2-\mu)e^{ikx}dk.
\end{equation}
All standard definitions, relations and frequency expansion take
place for $\Phi_{\alpha}(k,\mu)$, but the relation
$k^0_{\mu}=\sqrt{\bar{k}^2+\mu}$ defines smeared (fuzzy)
mass-shell due to random $\mu$.

The expressions (\ref{E:2}) and (\ref{E:3}) define the model
"free" unstable field, which really is some effective field. This
field is formed by interaction of "bare" UP with it's decay
channels and includes nonperturbative self-energy contribution in
the resonant region. Such an interaction leads to the spreading
(smearing) of mass from $\rho^{st}(\mu)=\delta(\mu-M^2)$ for the
bare particles to some smooth density function
$\rho(\mu)=|\omega(\mu)|^2$ with mean value $\bar{\mu}\approx M^2$
and $\sigma_{\mu}\approx \Gamma$. So, the UP is characterized in
the discussed model by the weight function $\omega(\mu)$ or
probability density $\rho(\mu)$ with parameters $M$ and $\Gamma$
(or real and imaginary parts of pole). A similar approach has been
discussed by Matthews and Salam in Ref. \cite{8}.

The commutative relations for a model operators have an additional
$\delta$-function:
\begin{equation}\label{E:6}
 [\dot{\Phi}^{-}_{\alpha}(\bar{k},\mu),\,\Phi^{+}_{\beta}(\bar{q},\mu^{'})]_{\pm}
 =\delta(\mu-\mu^{'}) \delta(\bar{k}-\bar{q})\delta_{\alpha\beta},
\end{equation}
where subscripts $\pm$ correspond to the fermion and boson fields.
The presence of $\delta(\mu-\mu^{'})$ in Eq.(\ref{E:6}) means an
assumption - the acts of creations and annihilations of particles
with various $\mu$ (random mass square) don't interfere. So, the
parameter $\mu$ has the status of physically distinguishable value
as random $m^2$. This assumption directly follows from the
interpretation of $q^2$ in Eq. (\ref{E:1}) as random parameter
$\mu$. By integrating both side of Eq.(\ref{E:6}) with weights
$\omega^{*}(\mu)\omega(\mu^{'})$ one can get standard commutative
relations
\begin{equation}\label{E:7}
 [\dot{\Phi}^-_{\alpha}(\bar{k}),\Phi^+_\beta(\bar{q})]_{\pm}=\delta(\bar{k}-\bar{q})
 \delta_{\alpha\beta}\,,
\end{equation}
where $\Phi^{\pm}_{\alpha}(\bar{k})$ is full operator field
function in momentum representation:
\begin{equation}\label{E:8}
 \Phi^{\pm}_{\alpha}(\bar{k})=\int\Phi^{\pm}_{\alpha}(\bar{k},\mu)\omega(\mu)d\mu\,.
\end{equation}
It should be noted that Eq.(\ref{E:7}) follows from Eq.(\ref{E:6})
when $\int|\omega(\mu)|^2d\mu=1$.

The expressions (\ref{E:2}) and (\ref{E:6}) are the principal
elements of the discussed model. The weight function $\omega(\mu)$
in Eq.(\ref{E:2}) (or $\rho(\mu)$) is full characteristic of UP
and the relations (\ref{E:6}) define the structure of the model
amplitude and of the transition probability (section 3). The
probability density $\rho(\mu)$ will be defined in the fourth
section by matching the model propagator to renormalized one.

With help of traditional method one can get from Eqs.(\ref{E:2}),
(\ref{E:4}) and (\ref{E:6}) the expression for the unstable scalar
Green function \cite{3}:
\begin{equation}\label{E:9}
 \langle 0|T(\phi(x),\phi(y))|0\rangle=D(x-y)=\int
 D(x-y,\mu)\rho(\mu)d\mu\,.
\end{equation}
In Eq.(\ref{E:9}) $D(x,\mu)$ is defined in the standard way for
the scalar field with $m^2=\mu$ and describes UP in an
intermediate state:
\begin{equation}\label{E:10}
 D(x,\mu)=\frac{i}{(2\pi)^4}\int\frac{e^{-ikx}}{k^2-\mu+i\epsilon}dk\,.
\end{equation}
The right side of the Eq.(\ref{E:9}) is Lehmann-like spectral (on
$\mu$) representation of the scalar Green function, which
describes the propagation of scalar UP. Taking into account the
connection between scalar and vector Green functions, we can get
the Green function of the vector unstable field
\begin{equation}\label{E:11}
 D_{mn}(x,\mu)=-(g_{mn}+\frac{1}{\mu}\frac{\partial^2}{\partial
 x^n\partial x^m})D(x,\mu)=\frac{-i}{(2\pi)^4}\int\frac{g_{mn}-k_m
 k_n/\mu}{k^2-\mu+i\epsilon}e^{-ikx}dk\,.
\end{equation}
Analogously Green function of the spinor unstable field:
\begin{equation}\label{E:12}
 \Hat{D}(x,\mu)=(i\hat{\partial}+\sqrt{\mu})D(x,\mu)=\frac{i}{(2\pi)^4}
 \int\frac{\hat{k}+\sqrt{\mu}}{k^2-\mu+i\epsilon}e^{-ikx}dk\,,
\end{equation}
where $\hat{k}=k_i\gamma^i$. These Green functions in momentum
representation have a convolution structure:
\begin{equation}\label{E:13}
 D_{mn}(k)=\int D_{mn}(k,\mu)\rho(\mu)d\mu\,,\,\,\,\,\,
 \Hat{D}(k)=\int \Hat{D}(k,\mu)\rho(\mu)d\mu\,.
\end{equation}

\section{Model amplitude and the convolution formula for a decay
rate}

In this section we consider the model amplitude for the simplest
processes with UP in a final state and get the CF (\ref{E:1}) as
direct consequence of the model. The expression for a scalar
operator field \cite{3}:
\begin{equation}\label{E:14}
 \phi^{\pm}(x)=\frac{1}{(2\pi)^{3/2}}\int\omega(\mu)d\mu\int\frac{a^{\pm}(\bar{q},\mu)}
 {\sqrt{2q^0_{\mu}}}e^{\pm iqx}d\bar{q}\,,
\end{equation}
where $q^0_{\mu}=\sqrt{\bar{q}^2+\mu}$ and $a^{\pm}(\bar{q},\mu)$
are creation or annihilation operators of UP with momentum $q$ and
mass square $m^2=\mu$. Taking into account Eq.(\ref{E:6}) we can
get:
\begin{equation}\label{E:15}
 [\dot{a}^{-}(\bar{k},\mu),\phi^{+}(x)]_{-};\, [\phi^{-}(x), \dot{a}^{+}(\bar{k},\mu)]_{-}
 =\frac{\omega(\mu)}{(2\pi)^{3/2}\sqrt{2k^0_{\mu}}}e^{\pm ikx}\,,
 \,\,\,k^0_{\mu}=\sqrt{\bar{k}^2+\mu}\,.
\end{equation}
The expressions (\ref{E:15}) differ from standard ones by the
factor $\omega(\mu)$ only. From this result it follows that, if
$\dot{a}^{+}(k,\mu)|0\rangle$ and $\langle0|\dot{a}^{-}(k,\mu)$
define UP with a mass $m^2=\mu$ and a momentum $k$ in the initial
and final states, then the amplitude for the decay of type
$\Phi\rightarrow\phi\phi_1$ has the form:
\begin{equation}\label{E:16}
 A(k,\mu)=\omega(\mu)A^{st}(k,\mu)\,,
\end{equation}
where $A^{st}(k,\mu)$ is amplitude in a stable particle
approximation when $m^2=\mu$. This amplitude is calculated in a
standard way and can include high corrections. Moreover, it can be
effective amplitude for the processes with hadron participation
\cite{3,5}.

To define the transition probability of the process
$\Phi\rightarrow\phi\phi_1$, where $\phi$ is UP with a large
width, we should take into account the status of parameter $\mu$
as physically distinguishable value, which follows from
Eq.(\ref{E:6}). Thus, the amplitude at different $\mu$ don't
interfere and we have the convolution structure of differential
(on $k$) probability:
\begin{equation}\label{E:17}
 d\Gamma(k)=\int d\Gamma^{st}(k,\mu)|\omega(\mu)|^2d\mu\,.
\end{equation}
In Eq.(\ref{E:17}) the differential probability
$d\Gamma^{st}(k,\mu)$ is defined in the standard way (stable
particle approximation):
\begin{equation}\label{E:18}
 d\Gamma^{st}(k,\mu)=\frac{1}{2\pi}\delta(k_{\Phi}-k_{\phi}-k_1)|A^{st}
 (k,\mu)|^2d\bar{k}_{\phi}d\bar{k}_1\,,
\end{equation}
where $k=(k_{\Phi},k_{\phi},k_1)$ denotes the momenta of
particles. From Eqs.(\ref{E:17}) and (\ref{E:18}) it directly
follows the known convolution formula for a decay rate
\begin{equation}\label{E:19}
 \Gamma(m_{\Phi},m_1)=\int_{\mu_0}^{\mu_m}\Gamma^{st}(m_{\Phi},m_1;\mu)\rho(\mu)d\mu\,,
\end{equation}
where $\rho(\mu)=|\omega(\mu)|^2$ and $\mu_0,\mu_m$ are defined in
Refs. \cite{5,7} as threshold and maximal invariant mass square of
unstable $\phi$.

An account of high corrections to the amplitude (\ref{E:16}) and,
hence, to Eq.(\ref{E:19}) keeps convolution form (\ref{E:19}).
This form can be destroyed by the interaction between the products
of UP ($\phi$) decay and initial $\Phi$ or final $\phi_1$ states.
The calculation in this case can be fulfilled in a standard way,
but UP in the intermediate state is described by the model
propagator. However, a calculation within the framework of
perturbative theory (PT) can not be applicable to the UP with
large width, that is to the short-living particle. In any case,
the applicability of PT, model approach or convolution method to
the discussed decays should be justified by experiment. The
correspondence of CM to the experimental data was demonstrated for
some processes \cite{3,5,6,7}, but this problem needs in more
detailed investigation. In this connection we should note the
analysis of higher-order corrections for processes with UP
\cite{4}. The separation between factorizable and non-factorizable
corrections make it possible to build the effective theory of UP
\cite{4}.

When there are two UP with large widths in a final state
$\Phi\rightarrow\phi_1\phi_2$, then in analogy with the previous
case one can get double convolution formula:
\begin{equation}\label{E:20}
 \Gamma(m_{\Phi})=\int\int\Gamma^{st}(m_{\Phi};\mu_1,\mu_2)\rho_1(\mu_1)\rho_2(\mu_2)d\mu_1
 d\mu_2\,.
\end{equation}
The derivation of CF for the cases when there is vector or spinor
UP in a final state can be done in analogy with the case of scalar
UP. However, in Eqs.(\ref{E:14}), (\ref{E:15}) and (\ref{E:16}) we
have a polarization vector $e_m(q)$ or spinor
$u^{\nu,\pm}_{\alpha}(q)$, where q is on fuzzy mass-shell. As a
result we get polarization matrix with $m^2=\mu$. For the vector
UP in a final state:
\begin{equation}\label{E:21}
 \sum_{e} e_m(q)e^{*}_n(q)=-g_{mn}+q_mq_n/\mu\,.
\end{equation}
For the spinor UP in a final state:
\begin{equation}\label{E:22}
 \sum_{\nu} u^{\nu,\pm}_{\alpha}(q)\bar{u}^{\nu,\mp}_{\beta}(q)=\frac{1}{2q^0_{\mu}}
 (\hat{q}\mp\sqrt{\mu})_{\alpha\beta}\,.
\end{equation}
In Eqs.(\ref{E:21}) and (\ref{E:22}) sum run over polarization and
$q^0_{\mu}=\sqrt{\bar{q}^2+\mu}$.

The formulae (\ref{E:19}) and (\ref{E:20}) describe FWE in full
analogy with the phenomenological convolution method \cite{5} and
with some cases of the decay-chain method \cite{5,7}. Thus, we
consider the quantum field basis for CM, which takes into account
the fundamental uncertainty principle and is in good agreement
with experimental date on some decays.  To evaluate FWE for the
case, when UP is in an initial state, we must account the process
of UP generation. When UP is in an intermediate state, then the
description of FWE is equivalent to the traditional one, but the
model propagators are determined by Eqs.(\ref{E:9}) -
(\ref{E:13}).

\section{Determination of $\rho(\mu)$ from renormalized propagator}

The possibility of $\rho(\mu)$-determination directly follows from
the connection of the decay-chain method (DCM) and convolution
method \cite{7}. As was shawn in Ref. \cite{7}, this connection
leads to the convolution formula (\ref{E:1}), where in accordance
with uncertainty principle $q^2$ is interpreted as smeared mass
square parameter $\mu$, which distribution is described by the
expression:
\begin{equation}\label{E:23}
 \rho(\mu)=\frac{1}{\pi}\frac{\sqrt{\mu}\,\Gamma(\mu)}{|P(\mu)|^2}\,.
\end{equation}
In Eq.(\ref{E:23}) $\Gamma(\mu)$ is $\mu$-dependent full width and
$P(\mu)^{-1}$ is propagator's denomenator. It should be noted,
that the convolution structure of Eq.(\ref{E:1}) and universal
structure of Eq(\ref{E:23}) don't depend on the definition of
$P(\mu)$. It has a complex pole structure $\mu-\mu_R$ and can be
approximated by the Breit-Wigner $\mu-M^2+iM\Gamma(\mu)$ \cite{5}
or another phenomenological approximation. The expression
(\ref{E:23}) is very simple and convenient in practical
calculations of decay rate, where the error of approximation is
small.

Here we'll consider the definition of $\rho(\mu)$ from the
matching model propagators to standard dressed ones \cite{3}. This
consideration is rather methodological than practical and
demonstrates the connection between model and traditional
descriptions. Let us associate the model propagator of scalar
unstable field (\ref{E:9}) with standard one:
\begin{equation}\label{E:24}
 \int\frac{\rho(\mu)d\mu}{k^2-\mu+i\epsilon}\longleftrightarrow
 \frac{1}{k^2-m^2_0-\Pi(k^2)}\,,
\end{equation}
where $\Pi(k^2)$ is conventional self-energy of scalar field. With
help of an analytical continuation of the expressions (\ref{E:24})
on complex plane $k^2\rightarrow k^2\pm i\epsilon$ and
prescription \cite{9}:
\begin{equation}\label{E:25}
 \Pi(k^2\pm i\epsilon)=Re\Pi(k^2)\mp iIm\Pi(k^2)
\end{equation}
the conformity (\ref{E:24}) can be represented by the equality
\begin{equation}\label{E:26}
 \int_{0}^{\infty}\frac{\rho(\mu}{k^2-\mu\pm
 i\epsilon}d\mu=\frac{1}{k^2-m^2(k^2)\pm
 iIm\Pi(k^2)}\,,
\end{equation}
where $m^2(k^2)=m^2_0+Re\Pi(k^2)$. With account of round pole
rules and $d\mu=d(\mu\mp i\epsilon), \rho(\mu\mp
i\epsilon)=\rho(\mu)\mp O(i\epsilon)$ two Eqs.(\ref{E:26}) can be
combine into the equality ($\mu\pm i\epsilon\rightarrow z)$:
\begin{equation}\label{E:27}
 \oint\frac{\rho(z)}{z-k^2}dz=\frac{1}{k^2-m^2(k^2)-iIm\Pi(k^2)}-
 \frac{1}{k^2-m^2(k^2)+iIm\Pi(k^2)}\,.
\end{equation}
The left side of Eq.(\ref{E:27}) is Cauchy integral, which equal
to $2\pi i\rho(k^2)$ and after a change $k^2\rightarrow\mu$ in the
final expression for $\rho$ we have:
\begin{equation}\label{E:28}
 \rho(\mu)=\frac{1}{\pi}\,\frac{Im\Pi(\mu)}{[\mu-m^2(\mu)]^2+[Im\Pi(\mu)]^2}\,.
\end{equation}
The expression (\ref{E:28}) for $\rho(k^2)$ in Breit-Wigner
approximation is usually exploited within the framework of
convolution method. From Eq.(\ref{E:28}) and definition
$\rho(\mu)=|\omega(\mu)|^2$ it follows:
\begin{equation}\label{E:29}
 \omega(\mu)=\frac{1}{\sqrt{\pi}}\,\frac{\sqrt{Im\Pi(\mu)}}{\mu-m^2(\mu)\pm
 iIm\Pi(\mu)}\,.
\end{equation}
The ambiguity of sign in (\ref{E:29}) is not essential because the
expression $|\omega(\mu)|^2$ only enters into the physical values.
In the parametrization $Im\Pi(\mu)=\sqrt{\mu}\,\Gamma(\mu)$ we
have relativistic Breit-Wigner $\omega(\mu)$ and Lorentzian
$\rho(\mu)$, which coincides with the expression (\ref{E:23}) for
renormalized  $P(q^2)$. Inserting the expression (\ref{E:28}) into
the left side of Eq.(\ref{E:24}) one can check with help of Cauchy
method the self-consistency of Eqs.(\ref{E:24}) and (\ref{E:28}).

Thus, we have put the correspondence between the model \cite{2} -
\cite{6} and some effective theory of UP with renormalized
propagator of scalar UP. To establish such a correspondence for
the vector UP we insert $\rho(\mu)$ into the model propagator
(\ref{E:13}) with $D_{mn}(k,\mu)$, defined by (\ref{E:11}) for
vector unstable field:
\begin{align}\label{E:30}
 \int_{0}^{\infty}\frac{-g_{mn}+k_m k_n
 /\mu}{k^2-\mu+i\epsilon}\,\frac{1}{\pi}\,\frac{Im\Pi(\mu)}{[\mu-m^2(\mu)]^2+
 [Im\Pi(\mu)]^2}\,d\mu=\\ \notag
 \frac{1}{2i\pi}\int_{0}^{\infty}\frac{-g_{mn}+k_m k_n
 /\mu}{k^2-\mu+i\epsilon}[\frac{1}{\mu-m^2(\mu)-i
 Im\Pi(\mu)}-\frac{1}{\mu-m^2(\mu)+iIm\Pi(\mu)}]d\mu\,.
\end{align}
With help of Eq.(\ref{E:25}) and above used method we can
represent the second part of Eq.(\ref{E:30}) in the
form($\mu\rightarrow z=\mu\pm i\epsilon$):
\begin{equation}\label{E:31}
 \frac{1}{2i\pi}\oint\frac{dz}{z-k^2}\,\,\frac{-g_{mn}+k_m k_n
 /z}{z-m^2(z)-iIm\Pi(z)}=\frac{-g_{mn}+k_m
 k_n/k^2}{k^2-m^2(k^2)-iIm\Pi(k^2)}\,.
\end{equation}
The right side of Eq.(\ref{E:31}) coincides with the expression
for propagator of vector UP, which leads to the convolution
formula (\ref{E:1}) in the decay-chain method \cite{7}. The
numerator of this effective propagator coincides with
$\eta_{mn}(k)$, which was used in \cite{7}. In Eqs.(\ref{E:30})
and (\ref{E:31}) the value $\Pi(k^2)$ is defined for vector field
as transverse part of polarization matrix \cite{1}. The
calculations of $\Pi(k^2)$ in effective theory (unstable hadrons)
or in gauge theory (Z,W-bosons) can run into some difficulties. In
the first case loop calculation can be ambiguous and we should use
traditional Breit-Wigner approximation $m^2(\mu)\approx M^2$ and
$Im\Pi(\mu)\approx\mu\Gamma(\mu)$. To escape the gauge-dependence
in the second case we can use pole definitions of mass and width
\cite{1}.

The description of $\rho(\mu)$ by the universal function
(\ref{E:28}) for scalar and vector fields can be justified by the
general structure of parametrization for bosons:
\begin{equation}\label{E:32}
 m^2(q^2)=m^2_0+Re\Pi(q^2),\,Im\Pi(q^2)=q\Gamma(q^2)\,.
\end{equation}
In the case of unstable fermion we have another parametrization
scheme:
\begin{equation}\label{E:33}
 m(q^2)=m_0+Re\Sigma(q^2),\,Im\Sigma(q^2)=\Gamma(q^2)\,.
\end{equation}
So, we need in additional analysis to define fermion function
$\rho(\mu)$. If we choose for fermion UP the universal density
function (\ref{E:23}), which follows from convolution method
\cite{7}, then we must do exchange $Im\Pi(\mu)\rightarrow
\sqrt{\mu}Im\Sigma(\mu)$ in the Eq.(\ref{E:28}). Inserting the
result into Eq.(\ref{E:13}) with $\hat{D}(x,\mu)$, defined by
Eq.(\ref{E:12}), we can get the correspondence between the model
propagator of fermion unstable field and the effective theory one:
\begin{equation}\label{E:34}
 \int\frac{\hat{k}+\sqrt{\mu}}{k^2-\mu+i\epsilon}\,\rho(\mu)d\mu\longrightarrow
 \frac{\hat{k}+k}{k^2-m^2(k^2)-ik\Sigma(k^2)}\,,
\end{equation}
where $k=\sqrt{(kk)}$.  The numerator of the right side of
Eq.(\ref{E:34}) coincides with the expression $\hat{\eta}$
\cite{7}.

The transitions (\ref{E:24}), (\ref{E:31}) and (\ref{E:34})
establish the correspondence between the discussed model and some
effective theory of UP within the framework of traditional QFT
approach. These transitions follow from the determination of
$\rho(\mu)$, that is from the accounting of interaction, which
forms the wave packet (\ref{E:2}) and mass-smearing. The most
important feature of the effective theory, chosen in such a way,
is the possibility to connect the decay-chain method and
convolution method within the framework of this theory \cite{7}.
So, we have some self-consistency of the discussed model,
effective theory, convolution and decay-chain method. However, due
to some difficulties, which arise in traditional approach, the
search of alternative $\rho(\mu)$ - definition is actual now.

\section{Conclusion}

The finite width effects in the processes with participation of UP
can be described by renormalized propagator, decay-chain method,
convolution method and effective theory of UP. The convolution
formula is convenient instrument for calculations of decay rate
and gives the results in accordance with experiment. In this paper
we have considered the model of UP with a random mass and derived
the convolution formula as a direct consequence of the model. The
model operator function and Lagrangian have a convolution
structure, which describes mass-smearing in accordance with
uncertainty principle.

The principal element of suggested model is probability density
function $\rho(\mu)$, which describes the main properties of UP.
Traditional description of UP in the intermediate state by
resonance line with complex pole (or by dressed propagator with
mass and width as parameters) corresponds to the model description
of UP in arbitrary state by function $\rho(\mu)$ with the same
parameters. We have considered the determination of $\rho(\mu)$
from DCM and by matching the model propagator to renormalized one.
This approach is equivalent to the convolution method or truncated
decay-chain method.

The second $\rho(\mu)$ - determination has some restrictions,
caused by propagator renormalization peculiarities. The question
arises, also, concern the possibility to describe mass-smearing of
bosons and fermions by the universal function $\rho(\mu)$.
Moreover, as the mass-smearing effect follows from the fundamental
uncertainty principle, then the search of $\rho(\mu)$ from the
first principles is reasonable. It should be noted also, that the
model erases a difference between the real and virtual states of
UP at peak region.

\end{document}